\documentclass[english,two column, showpacs]{revtex4}
\usepackage[T1]{fontenc}
\usepackage[latin1]{inputenc}
\usepackage{graphicx}
\usepackage{amsmath}
\usepackage{amssymb}
\usepackage{amsfonts}
\def\a{\alpha}
\def\b{\beta}
\def\g{\gamma}
\def\e{\varepsilon}
\def\d{\delta}

\def\l{\lambda}
\def\m{\mu}

\def\t{\tau}

\def\r{\rho}
\def\s{\sigma}

\def\S{\Sigma}
\def\G{\Gamma}
\def\D{\Delta}

\def\ua{\uparrow}
\def\da{\downarrow}
\def\pd{\partial}
\def\bk{{\bf k}}
\def\bK{{\bf K}}

\def\bp{{\bf p}}
\def\bn{{\bf n}}

\def\be{\begin{equation}}
\def\ee{\end{equation}}
\def\bea{\begin{eqnarray}}
\def\eea{\end{eqnarray}}
\def\nn{\nonumber}
\def\lb{\label}

\begin{document}

\title{Specifics of impurity effects in ferropnictide superconductors}
\author{Y.G. Pogorelov,$^1$ M.C. Santos,$^2$ V.M. Loktev$^3$}
\affiliation{$^1$IFIMUP-IN, Departamento de F\'{\i}sica, Universidade do Porto, Porto, Portugal,
\\ $^2$Departamento de F\'{\i}sica, Universidade de Coimbra, R. Larga, Coimbra, 3004-535,
Portugal, \\$^3$ Bogolyubov Institute for Theoretical Physics, NAN of
Ukraine, 14b Metrologichna str., 03143 Kiev, Ukraine}

\begin{abstract}
Effects of impurities and disorder on quasiparticle spectrum in superconducting iron
pnictides are considered. Possibility for occurrence of localized energy levels due to
impurities within the superconducting gap and the related modification of band structure
and of superconducting order parameter are discussed. The evolution of superconducting
state with impurity doping is traced.
\end{abstract}

\maketitle\

\section{Introduction}
The recent discovery of superconductivity (SC) with rather high critical temperature in the
family of doped ferropnictide compounds \cite{kamihara1, kamihara2}, has motivated a great
interest to these materials. Unlike the extensively studied cuprate family \cite{gins}, that
present insulating properties in their initial undoped state, the undoped LaOFeAs compound
is a semimetal. The scanning tunnel microscopy  (STM) study \cite{zhou} established that
this material has a layered structure, where the relevant for SC layer is FeAs with a 2D
square lattice by Fe atoms and with As atoms located out of plane, above or below the centers
of square cells (Fig. \ref{fig1}). Its electronic structure, relevant for constructing
microscopic SC models, have been explored with high-resolution angle-resolved photoemission
spectroscopy (ARPES) techniques \cite{ding, kondo}. Their results indicate the multiple
connected structure of Fermi surface, consisting of electron and hole pockets and absence
of nodes in both electron and hole gaps \cite{ding}, suggesting these systems to display
the so-called extended \emph{s}-wave SC order, changing its sign between electron and hole
segments
\cite{mazin}.

To study the band structure, the first principles numeric calculations are commonly used,
outlining the importance of Fe atomic \emph{d}-orbitals. The calculations show that SC in
these materials is associated with Fe atoms in the layer plane, represented in Fig. \ref{fig1}
by their orbitals and the related hopping amplitudes. The dominance of Fe atomic 3\emph{d}
orbitals in the density of states of LaOFeAs compound near its Fermi surface was demonstrated
by the local density approximation (LDA) calculations \cite{singh, haule, xu, mazin, cao, raghu}.
It was  then concluded that the multi-orbital effects are important for electronic excitation
spectrum in the SC state, causing formation of two gaps: by electron and hole pockets at the
Fermi surface.  To explain the observed SC properties, it is suggested that these
materials may reveal an unconventional pairing mechanism, beyond the common electron-phonon
scheme \cite{boeri, si}. In general, the total of 5 atomic orbitals for each iron in the
LaOFeAs compound can be involved, however the ways to reduce this basis are sought, in order
to simplify analytical and computational work. Some authors \cite{daghofer, tsai} have suggested
that it is sufficient to consider only the $d_{xz}$ and $d_{yz}$ orbitals. Thus, building such
minimal coupling model based on two orbitals, one is able to adjust the model parameters (energy
hopping and chemical potential) to obtain the Fermi surface with the same topology that found
in the first principles calculations of band structure.
\begin{figure}
\center \includegraphics[width=8cm]{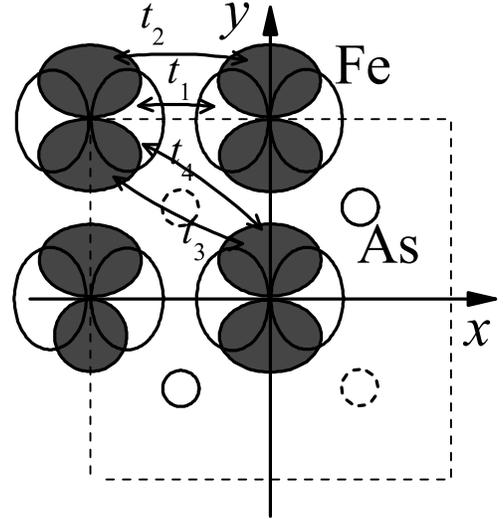}\\
\caption{Schematics of a FeAs layer in the LaoFeAs compound with $d_{xz}$ (white) and
$d_{yz}$ (dark) Fe orbitals and the Fe-Fe hopping parameters in the minimal coupling model.
Note that the hoppings between next near neighbors ($t_{3,4}$) are mediated by the As orbitals
(out of Fe plane).}
\lb{fig1}
\end{figure}

Having established the SC state parameters, an important class of problems can be considered
about the effects of disorder, in particular by impurities, on the system electronic properties,
and this issue has been also studied for doped ferropnictides. Alike the situation in doped
perovskite cuprates, here impurity centers can either result from the dopants, necessary to
form the very superconducting state, or from foreign atoms and other local defects in the
crystalline structure. Within the minimal coupling model, an interesting possibility for
localized impurity levels to appear within SC gaps in doped LaOFeAs was indicated, even for
the simplest, so-called isotopic (or non-magnetic) type of impurity perturbation \cite{dzhang, zhang}.
This finding marks an essential difference from the traditional SC systems with \emph{s}-wave
gap on a single-connected Fermi surface, were such perturbations are known not to produce
localized impurity states and thus to have no sizeable effect on SC transition temperature,
accordingly to the Anderson theorem \cite{and}. In presence of localized quasiparticle states
by isolated impurity centers, the next important issue is the possibility for collective
behavior of such states at high enough impurity concentrations. This possibility was studied
long ago for electronic quasiparticles in doped semiconducting systems \cite{ivanov} and
also for other types of quasiparticles in pnononic, magnonic, excitonic, etc. spectra under
impurities \cite{ilp}, establishing conditions for collective (including coherent) behavior
of impurity excitations with striking effects in observable properties of such systems. As
to the high-T$_c$ doped cuprates, it is known that their \emph{d}-wave symmetry of SC order
only permits existence of impurity resonances \cite{bal, pog}, not the true localization, and
hinders notable collective effects on their observable properties. As to our knowledgement,
no consistent study on collective impurity effects is know for the doped ferropnictide systems
up to the moment, and this defines the main emphasis of the present work. Namely, we shall
develop an analysis of these systems, using the Green function (GF) techniques, similar to those
for doped cuprate SC systems \cite{psl}, the minimal coupling model by two orbitals for
ferropnictide electronic structure, and the simplest isotopic type for impurity perturbation.
The structure of quasiparticle spectrum near in-gap impurity levels at finite impurity
concentrations, conditions for emergence of specific branches of collective excitations in
this region of the spectrum, and expected observable effects of such spectrum restructuring
will be discussed.

\section{Model Hamiltonian and Green functions}
For the minimal coupling model of Fig. \ref{fig1}, the hopping Hamiltonian $H_t$ is written
in the local orbital basis as:
\bea
 H_t & = & - \sum_{\bn,\s} \left[t_1\left(x_{\bn,\s}^\dagger x_{\bn + {\boldsymbol \d}_x,\s} +
  y_{\bn,\s}^\dagger y_{\bn + {\boldsymbol \d}_y,\s} + h. c.\right)\right.\nn\\
& + & t_2 \left(x_{\bn,\s}^\dagger x_{\bn + {\boldsymbol \d}_y,\s} +
y_{\bn,\s}^\dagger y_{\bn + {\boldsymbol \d}_x,\s} + h. c.\right)\nn\\
& + & t_3 \left(x_{\bn,\s}^\dagger x_{\bn + {\boldsymbol \d}_x + {\boldsymbol \d}_y,\s} +
x_{\bn,\s}^\dagger x_{\bn + {\boldsymbol \d}_x - {\boldsymbol \d}_y,\s} \right.\nn\\
& + & \left. y_{\bn,\s}^\dagger y_{\bn + {\boldsymbol \d}_x + {\boldsymbol \d}_y,\s} +
y_{\bn,\s}^\dagger y_{\bn + {\boldsymbol \d}_x - {\boldsymbol \d}_y,\s} + h. c.\right)\nn\\
& + & t_4 \left(x_{\bn,\s}^\dagger y_{\bn + {\boldsymbol \d}_x + {\boldsymbol \d}_y,\s} +
y_{\bn,\s}^\dagger x_{\bn + {\boldsymbol \d}_x + {\boldsymbol \d}_y,\s} \right.\nn\\
& - & \left.\left. x_{\bn,\s}^\dagger y_{\bn + {\boldsymbol \d}_x - {\boldsymbol \d}_y,\s}
- y_{\bn,\s}^\dagger x_{\bn + {\boldsymbol \d}_x - {\boldsymbol \d}_y,\s}
+ h. c.\right)\right].
\lb{eq1}
 \eea
where $x_{\bn,\s}$ and $y_{\bn,\s}$ are the Fermi operators for $d_{xz}$ and $d_{yz}$ Fe
orbitals with spin $\s$ on $\bn$ lattice site and the vectors ${\boldsymbol \d}_{x,y}$
point to its nearest neighbors in the square lattice. Passing to the operators of orbital
plane waves $x_{\bk,\s} = N^{-1/2} \sum_\bn{\rm e}^{i\bk\cdot\bn} x_{\bn,\s}$ (with the
number $N$ of lattice cells) and analogous $y_{\bk,\s}$, and defining an "orbital" 2-spinor
$\psi^\dagger(\bk,\s) = \left(x_{\bk,\s}, y_{\bk,\s}\right)$, one can expand the spinor
Hamiltonian in quasimomentum:
\be
 H_t = \sum_{\bk,\s} \psi^\dagger(\bk,\s)\hat h_t(\bk)\psi(k,\s).
 \lb{eq2}
  \ee
Here the 2$\times$2 matrix
\be
 \hat h_t(\bk) = \e_{+,\bk}\hat\s_0 + \e_{-,\bk}\hat\s_3 + \e_{xy,\bk}\hat\s_1
   \lb{eq3}
    \ee
includes the Pauli matrices $\hat\s_i$ and the energy functions
\[\e_{\pm,\bk} = \frac{\e_{x,\bk} \pm \e_{x,\bk}}2,\]
with
\bea
 \e_{x,\bk} & = & - 2t_1 \cos k_x - 2t_2 \cos k_y - 4t_3 \cos k_x \cos k_y,\nn\\
  \e_{y,\bk} & = & - 2t_1 \cos k_y - 2t_2 \cos k_x - 4t_3 \cos k_x \cos k_y,\nn\\
   \e_{xy,\bk} & = & - 4t_4 \sin k_x \sin k_y.
   \lb{eq4}
    \eea
\begin{figure}
\includegraphics[width=7cm]{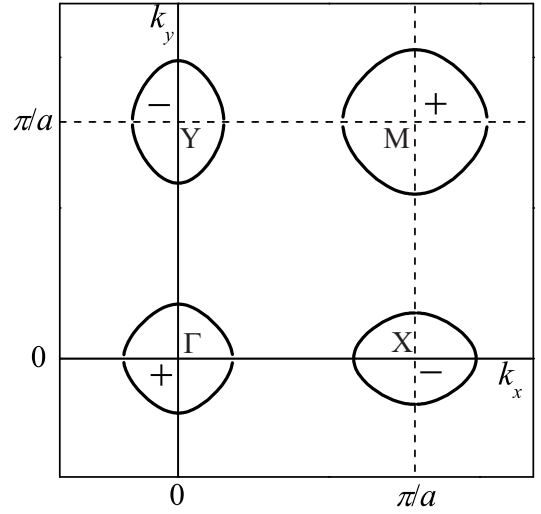}\\
\caption{Electron ($-$) and hole ($+$) segments of the Fermi surface in the normal state
of model system with electronic spectrum by Eq. \ref{eq5}. The center of first Brillouin
zone is displaced by $(\pi/2a,\pi/2a)$ to fully include all the segments around four
characteristic points $\G$, X, M, and Y in this zone. }
\lb{Fig2}
\end{figure}

An optimum fit for the calculated band structure within the minimum coupling model is attained
with the following set of hopping parameters (in $\left|t_1\right|$ units): $t_1 = -1.0,\,
t_2 = 1.3,\,t_3 = t_4 = -0.85$, and with the choice of the Fermi energy (chemical potential at
zero temperature)  $\e_{\rm F} = 1.45$ \cite{raghu}. The $\hat h_t$ matrix is diagonalized by
the standard unitary transformation:
\[\hat U(\bk) = \left(\begin{array}{cc}
                     \cos\theta_\bk/2 & -\sin\theta_\bk/2 \\
                     \sin\theta_\bk/2 & \cos\theta_\bk/2
                   \end{array}\right),\]
with $\theta_\bk = \arctan \left(\e_{xy,\bk}/\e_{-,\bk}\right)$, transforming from the
orbital to subband basis:
\be
 \hat h_b(\bk) = \hat U^\dagger(\bk)\hat h_t(\bk)\hat U(\bk) =
    \left(\begin{array}{cc}\e_{e,\bk} & 0 \\
                     0 & \e_{h,\bk}
                   \end{array}\right)
          \lb{eq5}
           \ee
The energy eigenvalues in Eq. \ref{eq4}:
\be
 \e_{h,e}(\bk) = \e_{+,\bk} \pm \sqrt{\e_{xy,\bk}^2 + \e_{-,\bk}^2},
 \lb{eq6}
  \ee
correspond to the two subbands in the normal state spectrum that respectively define
electron and hole pockets of the Fermi surface. There are two segments of each type,
defined by the equations $\e_{e,h}(\bk) = \m$, as shown in Fig. \ref{Fig2}. We note that
both functions $\cos\theta_\bk$ and $\sin\theta_\bk$ change their sign around these
segments, corresponding to their "azimuthal dependencies" around characteristic points of
the Brillouin zone (Fig. \ref{Fig2}), so that integrals of these functions with some
azimuthal-independent factors over the relevant vicinity of Fermi surface practically
vanish and are neglected beside such integrals of fully azimuthal-independent functions in
the analysis below.

The adequate basis for constructing the SC state is generated by the operators of electron
and hole subbands:
\bea
 \a_{\bk,\s} & = & x_{\bk,\s}\cos\theta_\bk/2 - y_{\bk,\s}\sin\theta_\bk/2,\nn\\
  \b_{\bk,\s} & = & y_{\bk,\s}\cos\theta_\bk/2 + x_{\bk,\s}\sin\theta_\bk/2,
   \lb{eq7}
    \eea
giving rise to the "multiband-Nambu" 4-spinors $\Psi_\bk^\dagger = \left(\a_{\bk,\ua}^\dagger,
\a_{-\bk,\da}, \b_{\bk,\ua}^\dagger,\b_{-\bk,\da}\right)$ and to a 4$\times$4 extension of
the Hamiltonian Eq. \ref{eq2} in the form:
\be
 H_s = \sum_{\bk,\s} \Psi_\bk^\dagger\hat h_s(\bk)\Psi_\bk,
 \lb{eq8}
  \ee
where the 4$\times$4 matrix
\[\hat h_s(\bk) = \hat h_b(\bk)\otimes\hat\t_3 + \D_{\bk}\hat\s_0\otimes\hat\t_1,\]
includes the Pauli matrices $\hat\t_i$ acting on the Nambu (particle-antiparticle) indices
in $\Psi$-spinors and $\hat h_b(\bk)$ is defined by Eq. \ref{eq5}. The simplified form for
the extended \emph{s}-wave SC order is realized with the definition of the gap function by
constant values, $\D_\bk = \D$ on the electron segments and $\D_\bk = -\D$ on the hole segments.

The electronic dynamics of this system is determined by the (Fourier transformed) GF 4$\times$4
matrices \cite{eco,ilp,psl}:
\be
 \hat G_{\bk,\bk'} = \langle\langle\Psi_\bk|\Psi_\bk^\dagger\rangle\rangle = i\int_{-\infty}^0
  dt {\rm e}^{i\e t/\hbar} \langle\{\Psi_\bk(t),\Psi_{\bk'}^\dagger(0)\}\rangle,
 \lb{eq9}
 \ee
whose energy argument $\e$ is understood as $\e -i0$ and $\langle\{A(t),B(0)\}\rangle$ is the
quantum statistical average with Hamiltonian $H$ of the anticommutator of Heisenberg operators.
From the equation of motion:
\be
 \e \hat G_{\bk,\bk'} = \hbar \d_{\bk,\bk'}\hat\s_0\otimes\t_0 + \langle\langle\left[\Psi_\bk,
  H\right]|\Psi_{\bk'}^\dagger\rangle\rangle,
   \lb{eq10}
    \ee
the explicit GF for the unperturbed SC system with the Hamiltonian $H_s$, Eq. \ref{eq7}, is
diagonal in quasimomentum, $\hat G_{\bk,\bk'} = \d_{\bk,\bk'}\hat G_\bk^0$ and
\bea
 \hat G_\bk^0 & = & \frac{\e\hat\t_0 + \e_e(\bk)
 \hat\t_3 + \D\hat\t_1}{2D_{e,\bk}}\otimes\hat\s_+\nn\\
 & + & \frac{\e\hat\t_0 + \e_h(\bk)\hat\t_3 - \D\hat\t_1}{2D_{h,\bk}}\otimes\hat\s_-,
  \lb{eq11}
   \eea
where $\hat\s_\pm = \left(\hat\s_0 \pm \hat\s_3\right)/2$ and the secular denominators
$D_{i,\bk} = \e^2 - \e_i^2(\bk) - \D^2$ for $i = e,h$. In what follows, we use the energy
reference to the Fermi level $\e_{\rm F}$ and approximate the segments of Fermi surface by
some circles of radius $k_i$ around the characteristic points $\bK_i$ in the Brillouin zone,
so that the dispersion laws $\e_j(\bk) = \e_{\rm F} + \xi_{j,\bk}$ permit to linearize the
quasiparticle dispersion close to the Fermi level as $\xi_{j,\bk} \approx \hbar v_j
\left(|\bk - \bK_j| - k_i\right)$. Generally, the Fermi wavenumbers $k_j$ and related Fermi
velocities $v_j$ for $j = e$ and $h$ can somewhat differ at a given choice of hopping parameters
and chemical potential, but, for simplicity, we shall neglect this difference and consider
their single values $k_j = k_{\rm F}$ and $v_j = v_{\rm F}$.

\section{Impurity perturbation and self-energy}

We pass to the impurity problem where the above Hamiltonian is added by the perturbation
terms due to non-magnetic impurities \cite{dzhang} on random sites $\bp$ in Fe square
lattice with an on-site energy shift $V$:
\be
 H_{imp} = V\sum_{\bp,\s} \left(x_{\bp,\s}^\dagger x_{\bp,\s} + y_{\bp,\s}^\dagger
  y_{\bp,\s}\right).
  \lb{eq12}
  \ee
Without loss of generality, the parameter $V$ can be taken positive, and for GF calculations,
this perturbation is suitably expressed in the multiband-Nambu basis:
\be
 H_{imp} = \frac 1 N \sum_{\bp,\bk,\bk'}{\rm e}^{i(\bk' - \bk)\cdot\bp}\Psi_\bk^\dagger
  \hat V_{\bk,\bk'}\Psi_{\bk'}.
   \lb{eq13}
    \ee
through the 4$\times$4 scattering matrix $\hat V_{\bk,\bk'} = V\hat U_\bk^\dagger
\hat U_{\bk'}\otimes\t_3$. Within the approach of Refs. \cite{ilp,psl}, the solution for
Eq. \ref{eq9} with the perturbed Hamiltonian $H_s + H_i$ can be obtained in different
forms, suitable for different types of states, band-like (extended) or localized.
All these forms result from the basic equation of motion:
\be
 \hat G_{\bk,\bk'} = \d_{\bk,\bk'}\hat G_\bk^0 +  \frac 1 N \sum_{\bp,\bk''}{\rm e}^{i(\bk''
  - \bk)\cdot\bp}\hat G_\bk^0\hat V_{\bk,\bk''}\hat G_{\bk'',\bk'},
   \lb{eq14}
    \ee
by specific routines of iterating this equation for the "scattered" GF's $\hat G_{\bk'',\bk'}$.

Thus, the algorithm, where the next iteration step \emph{never} applies to the scattered GF's
already present after previous steps, e.g. that with $\bk'' = \bk$ in Eq. \ref{eq14}, leads to
the so-called fully renormalized form, suitable for band-like states:
\be
 \hat G_\bk = \left[\left(\hat G_\bk^0\right)^{-1} - \hat \S_\bk\right]^{-1},
   \lb{eq15}
    \ee
where the self-energy matrix $\hat \S_\bk$ is expressed by the related group expansion (GE):
\be
\hat \S_\bk = c\hat T_\bk \left(1 + c \hat B_\bk + \dots\right).
 \lb{eq16}
  \ee
Here $c = \sum_\bp N^{-1}$ is the impurity concentration (per Fe site) and the T-matrix
results from all the multiple scatterings by a single impurity:
\bea
 \hat T_\bk & = & \hat V_{\bk,\bk} + \frac 1 N \sum_{\bk' \neq \bk}\hat V_{\bk,\bk'}
  \hat G_{\bk'}^0 \hat V_{\bk',\bk}\nn\\
& + & \frac 1 {N^2} \sum_{\bk'\neq \bk,\bk''\neq \bk,\bk'}\hat V_{\bk,\bk'}\hat G_{\bk'}^0
 \hat V_{\bk',\bk''} \hat G_{\bk''}^0 \hat V_{\bk'',\bk} + \dots.
  \lb{eq17}
   \eea
The next term to the unity in the brackets in Eq. \ref{eq14}:
\be
 \hat B_\bk = \sum_\bn \left(\hat A_{\bn}{\rm e}^{-i\bk\cdot\bn} + \hat A_{\bn}\hat A_{-\bn}
 \right)\left(1 - \hat A_{\bn}\hat A_{-\bn}\right)^{-1},
  \lb{eq18}
   \ee
describes the effects of indirect interactions in pairs of impurities, separated by vector
$\bn$, in terms of interaction matrices $\hat A_{\bn} = \hat T_\bk \sum_{\bk' \neq \bk}
{\rm e}^{i\bk'\cdot\bn}\hat G_{\bk'}$. Besides this restriction on summation, multiple sums
in the products like $\hat A_{\bn}\hat A_{-\bn}$ never contain coincident quasimomenta. Eq.
\ref{eq18} presents the first non-trivial GE term and the rest of its terms omitted in Eq.
\ref{eq14} correspond to the contributions from groups of three and more impurities \cite{ilp}.

An alternative iteration routine for Eq. \ref{eq13} applies it to \emph{all} the scattered
GF's, this results in the so-called non-renormalized form, suitable for localized states:
\be
 \hat G_\bk = \hat G_\bk^0 + \hat G_\bk^0\hat \S_\bk^0\hat G_\bk^0.
 \lb{eq19}
  \ee
Here the non-renormalized self-energy GE: $\hat \S_\bk^0 = c\hat T \left(1 + c\hat B_\bk^0 +
\dots\right)$, differs from the above renormalized one by absence of restrictions in
quasimomentum sums for  interaction matrices $\hat A_{\bn}^0 = \hat T_\bk \sum_{\bk'}
{\rm e}^{i\bk'\cdot\bn}\hat G_{\bk'}^0$ and their products.

At the first step, we shall restrict GE to the common T-matrix level, providing the conditions
for localized quasiparticle states with in-gap energy levels to appear at single impurities
\cite{tsai}, and will study certain (narrow) energy bands of specific collective states that
can be formed near these levels at finite impurity concentrations. At the next step, the
criteria for such collective states to really exist in the disordered SC system will follow
from the analysis of non-trivial GE terms. We notice that presence of renormalized GF's
$\hat G_{\bk'}$ in the above interaction matrices is just necessary for adequate treatment
of interaction effects over the in-gap bands.

\section{T-matrix and quasiparticle states}

The T-matrix, Eq. \ref{eq16}, is readily simplified taking into account that $\hat V_{\bk,\bk}
= V\hat\s_0\otimes\hat\t_3$ and introducing the integrated Green function matrix:
\[\hat G_0 =  \frac 1 N \sum_\bk\hat U_\bk\hat G_\bk^0 U_\bk^\dagger = \e\left[g_e(\e)\hat
\s_+ + g_h(\e)\hat\s_-\right]\otimes\hat\t_0.\]
This diagonal form follows directly from the aforementioned cancellation of the integrals
with $\cos\theta_\bk$ and $\sin\theta_\bk$ that appear in all the matrix elements of $\hat
U_\bk\hat G_\bk^0 U_\bk^\dagger$ except those proportional to $\e\hat\s_\pm\otimes\hat\t_0$.
Respectively, the functions $g_j(\e) = N^{-1}\sum_\bk D_{j,\bk}^{-1}$ for $j = e,h$ are
approximated near the Fermi level, $|\e - \e_{\rm F}|\lesssim \D$, as:
\be
 g_j(\e) \approx -\frac{\pi\r_j}{\sqrt{\D^2 - \e^2}}.
 \lb{eq20}
  \ee
Here $\r_j = m_j a^2/(2\pi\hbar^2)$ are the Fermi densities of states for respective subbands
(in parabolic approximation for their dispersion laws), and by the assumed identity of all
the segments of Fermi surface they can be also considered identical $\r_j = \r_{\rm F}$.
Omitted terms in Eq. \ref{eq16} are of higher orders in the small parameter $|\e|/\e_{\rm F}
\ll 1$.

Then the momentum independent T-matrix is explicitly written as
\be
 \hat T = \g^2\frac{\e - \e_0\hat\t_3}{\e^2 - \e_0^2},
 \lb{eq21}
  \ee
where $\e_0 = \D/\sqrt{1 + v^2}$ defines the in-gap impurity level \cite{tsai} through
the dimensionless impurity perturbation parameter $v = \pi\r_{\rm F}V$, and $\g^2 = v^2 V
\e_0^2/\D$ is the effective constant of coupling between localized and band quasiparticles.

\begin{figure}
\center  \includegraphics[width=9cm]{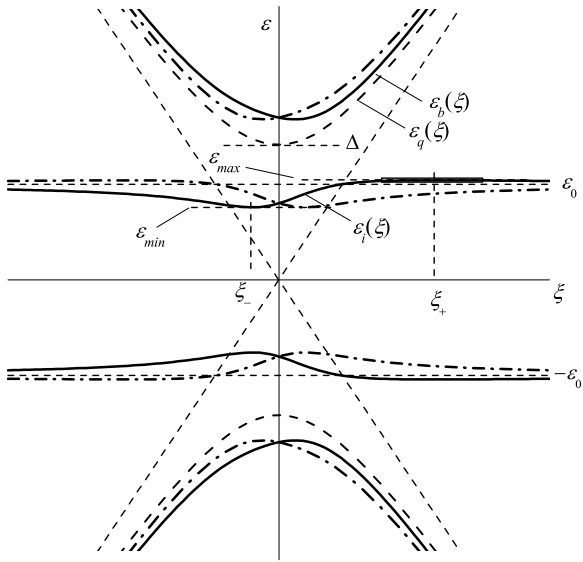}\\
\caption{Dispersion laws for band-like quasiparticles in the T-matrix approximation,
neglecting their finite lifetime, at a specific choice of impurity parameters $v = 1$,
$c = 0.1\D^2/\g^2$. Solid lines are for the bands near electron-like segments of Fermi
surface and dash-dotted lines for those near hole-like segments. The non-perturbed SC
quasiparticle bands and single-impurity localized levels are shown with dashed lines.
The narrow rectangle around the top of $\e_i$-band delimits the region in Fig. \ref{Fig5}.}
\label{Fig3}
\end{figure}

At finite $c$, using this T-matrix in Eq. \ref{eq14}, we obtain, from the condition $\det
\hat G_\bk^{-1} = 0$ \cite{eco}, the formal dispersion equation expressed through dispersion
of normal quasiparticles $\xi_\bk = \e_\bk - \e_{\rm F}$ (but in neglect of the energy level
width due to the effects of indirect interaction between impurities by higher GE terms):
\be
 D_\bk(\e) = \e^2 - \xi_\bk^2 - \D^2 - \frac{2 c\g^2\left(\e^2 - \e_0\xi_\bk\right)}{\e^2 -
 \e_0^2} = 0.
  \lb{eq22}
   \ee
Its solutions shown in Fig. \ref{Fig3} in function of the quasimomentum argument $\xi =
\xi_\bk$ display a peculiar multiband structure. First of all, it includes four modified
bands $\pm \e_b(\pm\xi)$, slightly shifted with respect to the unperturbed SC quasiparticle
bands $\pm \sqrt{\D^2 + \xi^2}$, accordingly to the basic function:
\be
 \e_b(\xi) \approx \sqrt{\D^2 + \xi^2} + c\g^2\frac{\D^2 + \xi^2 - \e_0\xi}{\sqrt{\D^2 +
  \xi^2}\left(\xi^2 + \xi_0^2\right)},
  \lb{eq23}
   \ee
with $\xi_0^2 = \D^2 - \e_0^2$.
It should be noted that these subbands for opposite signs of their argument $\xi$ in fact
refer to excitations around different segments (by electron and holes) of the Fermi surface,
but for clarity presented in Fig. \ref{Fig3} in the same $\xi$-reference. Besides these
$\e_b$ bands, there appear also four (narrow) in-gap bands $\pm \e_i(\pm\xi)$, generated
close to $\pm \e_0$ by finite concentration of impurities, accordingly to:
\be
 \e_i(\xi) \approx \e_0 + c\g^2\frac{\xi  - \e_0}{\xi^2 + \xi_0^2},
  \lb{eq24}
   \ee
As follows from Eq. \ref{eq21}, the $\e_j(\xi)$ band is
located between its extrema $\e_{max} = \e_0 + c\g^2\e_0/(\D + \e_0)$ at $\xi_+ = \e_0 + \D$
and $\e_{min} = \e_0 - c\g^2\e_0/(\D - \e_0)$ at $-\xi_- = \e_0 - \D$. The energy and momentum
shifts of the extremal points by Eqs. \ref{eq20}, \ref{eq21}, and Fig. \ref{Fig3} are specific
for the impurity effect on the multiband initial spectrum and they contrast with a simpler
situation for an impurity level near the edge of a single quasiparticle band \cite{ilp}.

All these spectrum bands would contribute to the overall density of states (DOS) by related
quasiparticles: $\r(\e) = (4\pi N)^{-1}{\rm Im \, Tr}\sum_\bk \hat G_\bk$. The more common
contributions here come from the $\e_b$ bands and they can be expressed through the
Bardeen-Cooper-Schrieffer (BCS) DOS in pure crystal \cite{tink}: $\r_{\rm BCS}(\e,\D) =
\r_{\rm F}\e/\sqrt{\e^2 - \D^2}$, as follows:
\be
 \r_b(\e) \approx \left(1 - \frac{c\g^2}{\e^2 - \e_0^2}\right)\r_{\rm BCS}\left(\e,\D_c\right),
 \lb{eq25}
 \ee
at $\e^2 \geq \D_c^2 = \D^2 + 2c\g^2\e_0^2/(\D^2 - \e_0^2)$. The first factor in the l.h.s.
of Eq. \ref{eq25} describes a certain reduction of the BCS DOS, especially when the energy
argument is close to the gap limits, and the shift of its gap argument is due to the
quantum-mechanical repulsion between the band and impurity levels.

\begin{figure}
\center  \includegraphics[width=9cm]{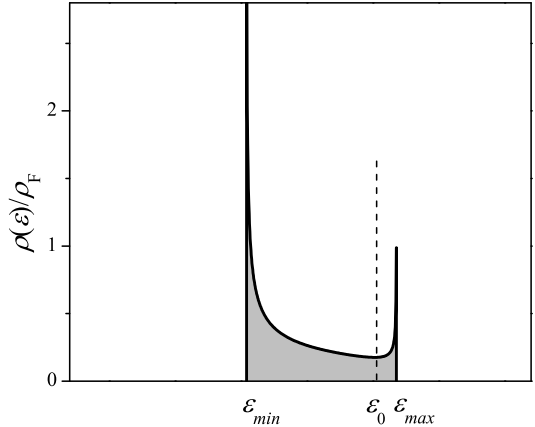}\\
\caption{Density of states in the narrow in-gap band near the impurity level $\e_0$
(dashed line) for the case by Fig. \ref{Fig3}.}
\label{Fig4}
\end{figure}
More peculiar is the contribution to DOS from the $\e_i$ bands, written as:
\be
 \r_i(\e) \approx \frac{\r_{\rm F}}v\frac{\e^2 - \e_0^2 - c\g^2}{\sqrt{\left(\e_{max}^2 -
 \e^2\right)\left(\e^2 - \e_{min}^2\right)}},
 \lb{eq26}
 \ee
at $\e_{min}^2 \leq \e^2 \leq \e_{max}^2$, and presented in Fig. \ref{Fig4}.

Both the effects of $\e_b$ band shifts and of $\e_i$ band formation can have important
repercussions in the physical behavior of the disordered SC system and they will be
considered below. But before this, we need to analyze the criteria for the considered
quasiparticles to really exist, especially in closeness to the limits of corresponding
bands.

\section{Group expansion and coherence criteria}
Let us now study the crossover from band to localized states near the limits of $\e_i$ bands,
say for definiteness, its upper limit $\e_{max}$. Supposing the actual energy $\e < \e_{max}$
to be within the range of band states, we use the fully renormalized self-energy matrix, Eq.
\ref{eq16}, up to the GE pair term, $c^2 \hat T \hat B_\bk$,  that will add a certain finite
imaginary part $\Gamma_i(\xi)$ to the dispersion law $\e = \e_i(\xi)$, Eq. \ref{eq23}. Then
the known Ioffe-Regel-Mott criterion \cite{ioffe, mott} for the state at this energy to be
really band-like (also called extended) is written as:
\be
 \e_{max} - \e \gg \Gamma_i(\e).
  \lb{eq27}
   \ee
To simplify calculation of the scalar function $\Gamma_i(\e)$, we fix the energy argument in
the numerators of T-matrix and interaction matrices at $\e = \e_0$, obtaining their forms:
\be
 \hat T(\e) \approx \frac{\g^2\e_0}{\e^2 - \e_0^2}\hat m_+,\quad \hat A_\bn(\e)
 \approx \hat T(\e)\frac{\e_0}{N} \sum_\bk \frac{{\rm e}^{i\bk\cdot\bn}}{D_\bk(\e)},
  \lb{eq28}
   \ee
\begin{figure}
\center \includegraphics[width=9cm]{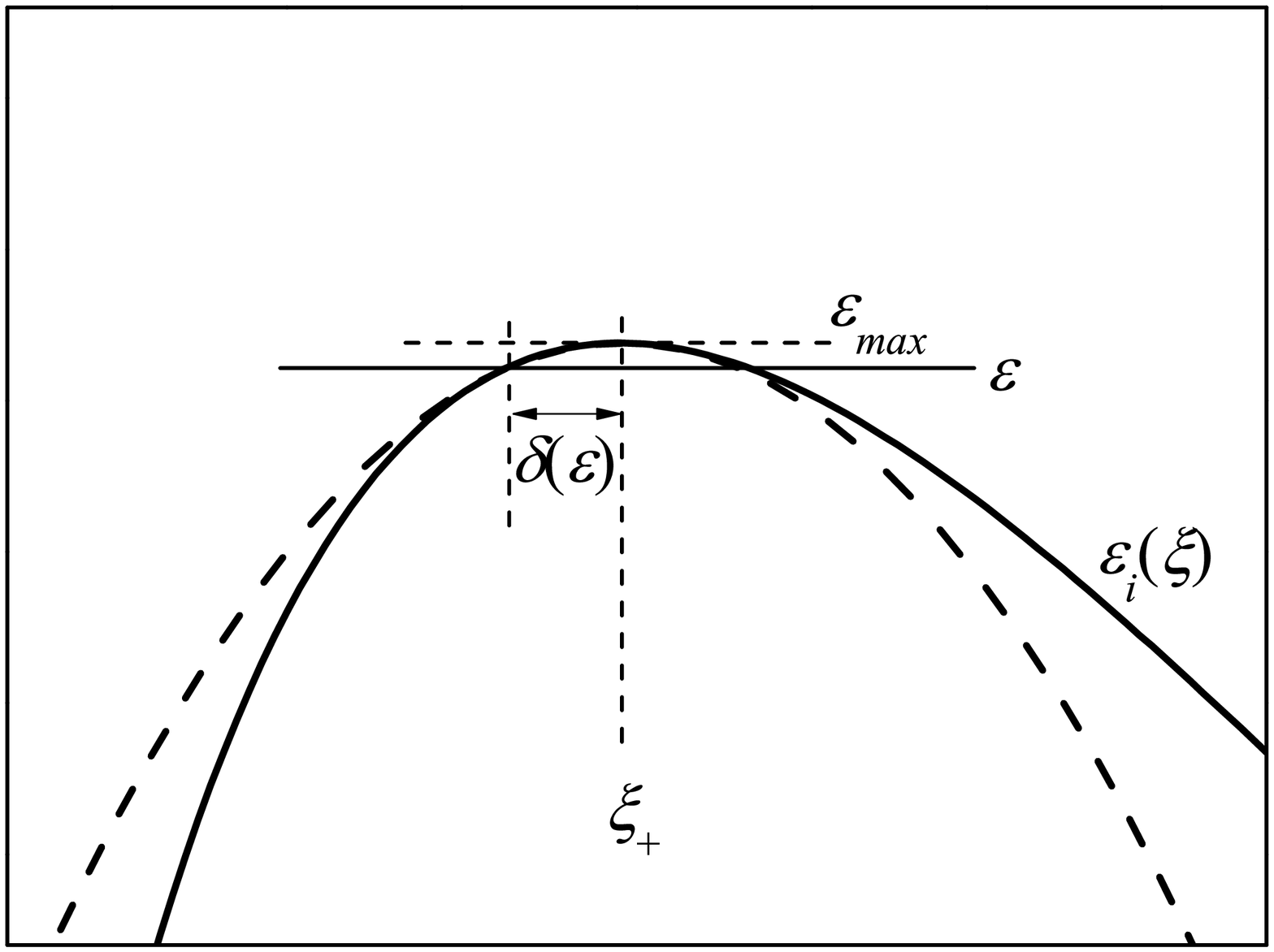}\\
  \caption{Parabolic approximation (dashed line) for the dispersion law near the top of
  impurity    band (solid line).}
    \lb{Fig5}
\end{figure}
both proportional to the matrix $\hat m_+ = \hat\s_0\otimes(\hat\t_0 + \hat\t_3)$ with
important multiplicative property: $\hat m_+^2 = 2\hat m_+$. The $\bk$-summation (integration)
in Eq. \ref{eq28} is suitable done in polar coordinates over the circular segments of Fermi
surface. Here the azimuthal integration only refers to the phase of numerator, resulting in
a zeroth order Bessel function: $\int_0^{2\pi}{\rm e}^{i x\cos\theta} d\theta = 2\pi J_0(x)$.
Since $x = n\left(k_{\rm F} + \xi/\hbar v_{\rm F}\right)$ is typically big, $x \gg 1$, the
asymptotical formula applies: $J_0(x) \approx \sqrt{2/(\pi x)}\cos(x - \pi/4)$. Then, for
radial integration in $\xi$ around the extremum point $\xi_+$, it is convenient to decompose
this function in the fast and slow oscillating factors: $J_0(x) \approx \sqrt{2/(\pi k_+ n)}
\cos(k_+ n - \pi/4)\cos[\left(\xi - \xi_+\right)n/\hbar v_{\rm F}]$ with the fast wavenumber
$k_+ = k_{\rm F} + \xi_+/\hbar v_{\rm F} \approx  k_{\rm F}$, and to write the denominator
in the parabolic approximation: $D_\xi(\e) \approx \left(\xi - \xi_+\right)^2 - \d^2(\e)$,
with $\d^2(\e) = 4\D\left(\D + \e_0\right)^2 \left(\e_{max} - \e\right)/(2c\g^2)$ (see Fig.
\ref{Fig5}). Thus, the interaction matrix $\hat A_\bn(\e) = A_n(\e)\hat m_+$ only depends
on the distance $n$ between impurities, and, for $\e$ close to $\e_{max}$, this dependence
can be expressed as:
\be
 A_r(\e) \approx \sqrt{\frac{r_\e}r}\sin k_\e r \cos k_{\rm F}r,
  \lb{eq29}
   \ee
where the length scales both for the monotonous decay:
\[r_\e = \frac{2\pi}{k_{\rm F}}\left[\frac{\e_0\r_{\rm F}\left(\D + \e_0\right)}
{c\d(\e)}\right]^2,\]
and for the sine factor: $k_\e^{-1} = \hbar v_{\rm F}/\d(\e)$, are much longer than
$k_{\rm F}^{-1}$ for the fast cosine. The latter fast oscillation is specific for the
interactions mediated by Fermi quasiparticles (like the known RKKY mechanism), unlike the
monotonous or slowly oscillating interactions between impurities in semiconductors or in
bosonic systems \cite{ilp}.
\begin{figure}
\center  \includegraphics[width=9cm]{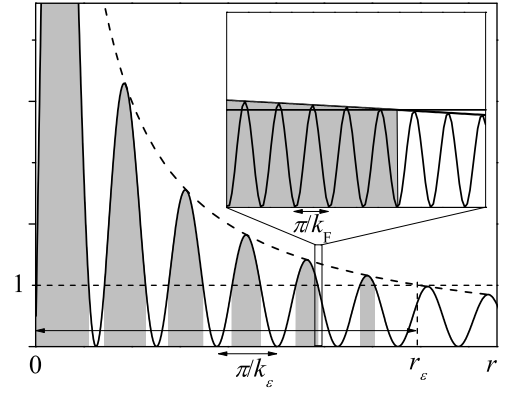}\\
  \caption{Interaction function $A_r^2(\e)$ by Eq. \ref{eq30} at the choice of parameters
  $\e_{max} - \e =   0.1$   and   $\D/\e_{\rm F} = 5\cdot 10^{-2}$ displays slow sine
  oscillations (solid line) and the monotonous envelope function (dashed   line). The shadowed
  intervals are those contributing to Im $B$, accordingly to   the   condition $(r_e/r)\sin^2
  k_\e r > 1$. Inset: the expansion of the rectangle in the main panel shows also fast
  oscillations by the cosine.}
  \label{Fig6}
\end{figure}
Now the calculation of $\G_i(\e)  = c^2 T(\e) {\rm Im} B(\e)$ mainly concerns the dominant
scalar part of the GE pair term:
\be
  B(\e) \approx \frac {2\pi}{a^2}\int_a^{r_\e}\frac{r \,dr}{1 - 4 A_r^2(\e)}
   \lb{eq30}
    \ee
(since the $\bk$-dependent term in Eq. \ref{eq18} turns to be negligible beside this).

The upper integration limit in Eq. \ref{eq31} corresponds to the condition that its
integrand only has poles for $r < r_\e$. In conformity with the slow and fast modes in the
function, Eq. \ref{eq30}, the integration is naturally divided in two stages. At the first
stage, integration over each $m$th period of fast cosine, around $r_m = 2\pi m/k_{\rm F}$,
is done setting constant the slow factors, $r \approx r_m$ and $\sin k_\e r \approx \sin
k_\e r_m$, and using the explicit formula:
\be
 {\rm Im} \int_{-\pi}^\pi \frac{dx}{1 - 4 A^2 \cos^2 x} =  {\rm Im}\frac {\pi}{\sqrt{1 -
   A^2}}.
   \lb{eq31}
    \ee
At the second stage, the summation of these results in $m$ is approximated by the integration
in the slow variable:
\bea
 &&\frac{\pi}{k_{\rm F}} {\rm Im} \sum_m \frac {r_m^{3/2}}{\sqrt{ r_m - r_\e\sin^2 k_\e
  r_m}}\nn\\
 &&\qquad \approx {\rm Im}\int_a^{r_\e}\frac{r^{3/2}dr}{\sqrt{r - r_\e\sin^2 k_\e r}}.
    \lb{eq32}
     \eea
The numerical calculation of the latter integral results in:
\be
 {\rm Im} B = \frac{r_\e^2}{a^2} f\left(k_\e r_\e\right),
  \lb{eq33}
   \ee
where the function $f(z)$ is zero for $z < z_0 \approx 1.3585$, and monotonously grows for
$z > z_0$, rapidly approaching the asymptotic constant value: $f_{as} \approx 1.1478$, for
$z \gg z_0$. Then the Ioffe-Regel-Mott criterion, Eq. \ref{eq27}, at $\e$ so close to
$\e_{max}$ that $k_\e r_\e \gg z_0$, is expressed as:
\be
 \e_{max} - \e \gg  \frac{c^2\g^2}{\e_{max} - \e_0}\frac{r_\e^2}{a^2},
  \lb{eq34}
   \ee
and this would result in a (concentration independent) estimate for the range of extended
states within the impurity band:
\be
 \e_{max} - \e \gg \G_0 = \frac{(v\e_0)^{3/2}}{ak_{\rm F}}\sqrt{\frac{2\pi \r_{\rm F}}{1 +
  v^2}},
   \lb{eq35}
    \ee
and its comparison with the full extension of this band, $\e_{max} - \e_{min} = c\g^2(1 +
v^2)/(v^2\D)$, would suggest possibility for such extended states to really exist if the
impurity concentration surpass the characteristic (small) value:
\be
 c \gg c_0 = \frac {\left(\pi\r_{\rm F}\e_0\right)^{3/2}}{a k_{\rm F}}\sqrt{\frac{2v}{1 + v^2}}.
  \lb{eq36}
   \ee
For typical values of $\r_{\rm F}^{-1} \sim 2$ eV, $a k_{\rm F} \sim 1$, and $\D \sim 10$
meV in LaOFeAs system \cite{ding, haule, schma}, and supposing a plausible impurity perturbation 
$v \sim 1$, we estimate $c_0 \approx 8\cdot 10^{-4}$, manifesting important impurity effects 
already at their very low content.

However, the r. h. s. of Eq. \ref{eq34} vanishes at $k_\e r_\e < z_0$, which occurs
beyond the vicinity of the band top:
\be
 \e_{max} - \e > \G_0\left(\frac{c_0}c\right)^3.
  \lb{eq37}
   \ee
Under the condition of Eq. \ref{eq36}, this vicinity is yet more narrow than $\G_0$ by Eq.
\ref{eq35}, defining the true, even wider, range of extended states.
\begin{figure}
\center  \includegraphics[width=9cm]{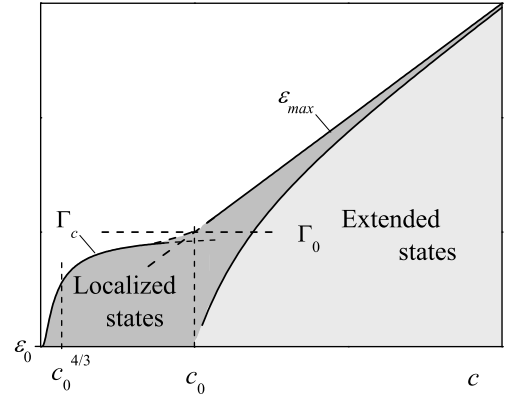}\\
  \caption{Structure of the energy spectrum near the impurity level in function of impurity
  concentration.}
  \lb{Fig7}
\end{figure}

Otherwise, for $c \ll c_0$, the impurity band does not exist, then we analyze the energy
range near the impurity level with the non-renormalized GE and write the approximate
criterion for its convergence as $c|B^0| \ll 1$. This calculation is done in a similar way
as before but replacing the interaction function, Eq. \ref{eq29}, by its non-renormalized
version:
\be
  A_r^0(\e) \approx \sqrt{R_\e/r}\, {\rm e}^{-r/r_0}\cos k_{\rm F}r,
  \lb{eq38}
   \ee
with $k_{\rm F}R_\e = 2\pi\left(\e_0/|\e - \e_0|\right)^2$ and $k_{\rm F}r_0 = 2\e_{\rm F}/
\xi_0$. Then the above GE convergence criterion is assured beyond the following vicinity of
impurity level:
\be
 |\e - \e_0| \gg \G_c = \G_0 \exp\left(-c_0^{4/3}/c\right),
 \lb{eq39}
  \ee
defining the range of its broadening due to inter-impurity interactions. The DOS function for
localized states can be only estimated by the order of magnitude within this range, but outside
is given by:
\bea
 \r_{loc}(\e) & \approx & \frac{c^2}{c_0^{4/3}|\e - \e_0|},\quad{\rm for}\quad\G_c \ll |\e -
  \e_0| \ll \G_0,\nn\\
   \r_{loc}(\e) & \approx  & \frac{c^2\e_0^4}{|\e - \e_0|^5},\quad{\rm for}\quad\G_0 \ll |\e -
    \e_0|.
    \lb{eq40}
     \eea
Notably, the total number of states near the impurity level is $\int \r_{loc}(\e)d\e \sim c$,
alike that of extended states in the impurity band by Eq. \ref{eq26}. The summary of evolution
of this area of quasiparticle spectrum in function of impurity concentration is shown in Fig.
\ref{Fig7}.

\section{Impurity effects on SC characteristics}
The above results on the quasiparticle spectrum in the disordered SC system can be immediately
used for calculation of impurity effects on its observable characteristics.

Thus the fundamental SC order parameter $\D$ is estimated from the modified gap equation:
\be
 \l^{-1} = \int_0^{\e_{\rm D}}\r(\e)d\e,
  \lb{eq41}
   \ee
where $\l = \r_{\rm F}V_{SC}$ is the (small) dimensionless SC pairing constant and the Debye
energy $\e_{\rm D}$ restricts the energy range of its action. In absence of impurities, $c =
0$, using the BCS DOS in this equation leads straightforwardly to the known result for its
non-perturbed value $\D_0$: $\l^{-1} = {\rm arcsinh}\left(\e_{\rm D}/\D_0\right)$ and thus to
$\D_0 \approx \e_{\rm D}{\rm e}^{-1/\l}$.

For finite $c$, the total DOS is combined from the contributions by the shifted main band,
$\r_b$, Eq. \ref{eq25}, and by the impurity band (or level) $\r_i$ (or $\r_{loc}$), Eqs. \ref{eq26}
(or \ref{eq40}). The latter contribution is $\sim c$, accordingly to the previous discussion,
defining a small correction beside $\l^{-1} \gg 1$. But a much stronger $c$-dependent correction
comes from the modified main band:
\bea
 \int_{\D_c}^{\e_{\rm D}}\r_b(\e)d\e & \approx & {\rm arcsinh}\frac{\e_{\rm D}}{\D_c}\nn\\
  & - & c\g^2\int_{\D_c}^{\e_{\rm D}}\frac{d\e}{\left(\e - \e_0\right)^2\sqrt{\D_c^2 - \e^2}}.\nn
  \eea
For ${\e_{\rm D}} \gg \D_c$, the last integral is well approximated by:
\[c\g^2\int_{\D_c}^\infty\frac{d\e}{\left(\e - \e_0\right)^2\sqrt{\D_c^2 - \e^2}} =
\frac{c\g^2}{\D_c^2}F\left(\frac{\D_c}{\e_0}\right),\]
with the function
\[F(z) = z\frac{\sqrt{z^2 - 1} + z \arccos(-1/z)}{\left(z^2 - 1\right)^{3/2}}.\]
Though this $F$ diverges at $z \to 1$, but actually its argument
\[\D_c/\e_0 = \sqrt{1 + v^2}\left(1 + c/c_1\right) \quad {\rm with}\quad
c_1 = \pi\r_{\rm F}\D/v,\]
is always above unity. Neglecting the small $\r_i$ contribution in Eq. \ref{eq41} and taking
account of the BCS relation $\l^{-1} = {\rm arcsinh}\left(\e_{\rm D}/\D_0\right)$, we express
the gap equation as:
\be
  {\rm arcsinh}\frac{\D_c - \D_0}{\D_0} \approx \frac {c v^2}{c_1\left(1 + v^2\right)}F
  \left(\D_c/\e_0\right).
  \lb{eq43}
   \ee
Its approximate solution for $c \ll c_1$, together with the relation: $\D_c/\D = 1 + c
/\left[c_1 \left(1 + v^2\right)\right]$, lead to the desired expression for the perturbed
SC order parameter $\D$:
\be
  \frac{\D}{\D_0} \approx 1 - \frac c{c_1}\frac{1 + v^2 F\left[\sqrt{1 + v^2}\left(1 +
  c/c_1\right)\right]}{1 + v^2},
  \lb{eq44}
   \ee
that rapidly decays with impurity concentration and would vanish at
\[c = c_1\frac{1 + v^2}{1 + v^2 F\left[\sqrt{1 + v^2}\left(1 + c/c_1\right)\right]}.\]
The latter equality defines in fact a certain equation for $c$ and its solution, e.g., for
the above choice of $v = 1$, is $c \approx 0.5c_1 \approx 6\cdot 10^{-3}$. However, such
concentrations would already correspond to the impurity band as wide as the gap itself;
this goes beyond the validity of the above derivation and needs a special treatment (to
be done elsewhere).

To study another important dependence, that of the SC transition temperature $T_c$ on
concentration $c$, one has, strictly speaking, to extend the above GF techniques for finite
temperatures, but a very simple estimate can be done, supposing that the BCS relation
$\D/T_c \approx 1.76$ still holds in the presence of impurities. Then the r.h.s. of Eq.
\ref{eq44} would also describe the decay of $T_c/T_{c0}$.

It is of interest to compare the present results with the known Abrikosov-Gor'kov solution
for BCS SC with paramagnetic impurities in the Born approximation \cite{deew, sriva}. In
that approximation, the only perturbation parameter is the (constant) quasiparticle lifetime
$\t$. In our framework, the $\t^{-1}$ can be related to ${\rm Im}\S(\e)$ at a proper choice
of energy, $\e \sim |\D - \e| \sim \D$. Then, in the self-consistent T-matrix approximation
\cite{psl}, we estimate $\t^{-1} \sim c\D/c_1$ which leads to the relation $\t T_c \sim
c_1/c$, reaching at $c \gtrsim c_1$ a qualitative agreement with the Abrikosov-Gor'kov
universal criterion for complete SC suppression $\t T_c < 0.567$ (though in our case
this criterion is not universal and depends yet on the perturbation parameter $v$).
\begin{figure}
  % Requires \usepackage{graphicx}
  \includegraphics[width=9cm]{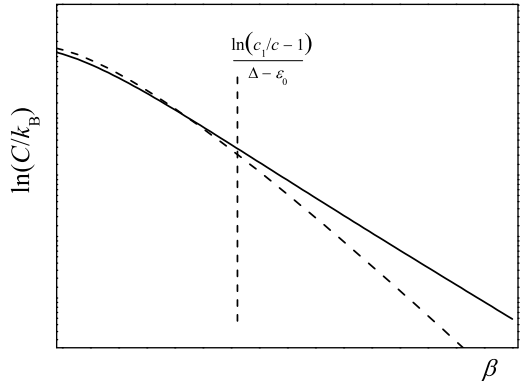}\\
  \caption{Temperature behavior of specific heat for a SC with impurities presents a
  crossover from $\b\D$ exponent (dashed line) to $\b\e_0$ at low enough temperature (high
  enough $\b = 1/k_{\rm B}T$).}\label{Fig8}
\end{figure}

Finally, a similar analysis can be applied for the impurity effect on the electronic
specific heat in the SC state, whose dependence on inverse temperature $\b = 1/k_{\rm B}T$
is represented as:
\be
 C(\b) = \frac \pd{\pd T} \int_0^\infty \frac{\r(\e)d\e}{{\rm e}^{\b\e} + 1},
 \lb{eq45}
 \ee
and naturally divided in two characteristic contributions, $C = C_i + C_b$, from $\r_i$
and $\r_b$ states:
\[C_i(\b) \approx k_{\rm B}c\left[\frac{\b\e_0}{2\cosh\left(\b\e_0/2\right)}
\right]^2,\]
and
\[C_b(\b) \approx  k_{\rm B}(c_1 - c)v\left(\b\D_c\right)^{3/2}\exp\left(-\b\D_c\right).\]
The resulting function $C(\b)$ deviates from the known low temperature behavior $C_0(\b)
\sim \exp(-\b\D)$ for non-perturbed SC system at $\b > \ln(c_1/c - 1)/(\D - \e_0)$, where
the characteristic exponent is changed to a slower $\sim \exp(-\b\e_0)$ as seen in Fig.
\ref{Fig8}.

The same approach can be used for calculation of other observable characteristics for SC
state under impurity effect, such as, e.g., differential conductivity for scanning tunneling
spectroscopy or absorption coefficient for far infrared radiation, though these issues are
beyond the scope of this work.

\section{Conclusions}
Resuming, the Green function analysis of quasiparticle spectra in an SC ferropnictide with
impurities permits to describe formation of impurity localized levels within SC gap and,
with growing impurity concentration, of specific band of extended quasiparticle states,
mainly supported by impurity centers. Explicit dispersion laws and densities of states are
obtained for modified main bands and impurity bands. Further specification of the nature of
all the states in different energy ranges within the SC gap is attained with analysis of group
expansions for self-energy matrix, resulting in criteria for crossovers between localized and
extended states. The developed spectral characteristics are applied for description of observable
impurity effects.

\end{document}